\documentstyle[aps,subeqnarray,prb,graphicx,multicol]{revtex}

\def\bs#1{\bbox{#1}}
\def\ve#1{\bs{\mathbf{#1}}}   
\newcommand{\be}{\begin{eqnarray}}
\newcommand{\ee}{\end{eqnarray}}

\begin{document}

\draft \title{Time averaged total force on a dipolar sphere in an
electromagnetic field}

\author{P. C. Chaumet and M. Nieto-Vesperinas}

\address{Instituto de Ciencia de Materiales de Madrid, Consejo
Superior de investigaciones Cientificas, Campus de Cantoblanco Madrid
28049, Spain}

\maketitle

\begin{abstract}        

We establish the time averaged total force on a subwavelength sized
particle in a time harmonic varying field. Our analysis is not
restrictive about the spatial dependence of the incident field. We
discuss the addition of the radiative reaction term in the
polarizability in order to correctly deal with the scattering
force. As a consequence and illustration, we assess the degree of
accuracy of several polarizability models previously established.
\end{abstract}

\begin{multicols}{2}

In the last years there has been an increase of interest in the
manipulation of small particles by means of the light Lorentz's
force. For spheres subwavelength radius the total force due to a light
wave is usually split into two parts from the use of the dipole
approximation (cf. Ref.~[\ref{gordon}]): a gradient force
$(\ve{p}.\ve{\nabla})\ve{E}$, essentially due to the particle induced
dipole moment $\ve{p}$ interacting with the electric field $\ve{E}$;
and a scattering and absorbing forces $\frac{1}{c}\dot{\ve{p}} \times
\ve{B}$, where $\ve{B}$ is the magnetic vector,
$\dot{\ve{p}}=\partial\ve{p}/\partial t$, and $c$ is the speed of
light in vacuum. It has been customary after Ref.~[\ref{gordon}] to
express the gradient force $\ve{F}_{\text{grad}}$ as (see
e.g. Ref.~[\ref{ashkin1}]): \be\label{forcem} \ve{F}_{\text{grad}}=
(1/2) \alpha_0 \ve{\nabla}\ve{E}^2,\ee where $\alpha_0$ is the
particle polarizability, satisfying the Clausius-Mossotti equation:
\be\label{polacm} \alpha_0=a^3\frac{\epsilon-1}{\epsilon+2},\ee $a$
being the particle radius and $\epsilon$ denoting the dielectric
permittivity. On the other hand, the absorbing and scattering forces
are written in the approximation of small spheres through the
absorbing ($C_{abs}$) and scattering ($C_{scat}$) cross sections as:
\be\label{fscatt} \ve{F}=
\frac{|E|^2}{(8\pi)}(C_{abs}+C_{scat})\frac{\ve{k}}{k},\ee where
$\ve{k}$ represents the light vector ($k=|\ve{k}|$).  On using the
expression of these cross sections in the dipole approximation, just
the first term of their Taylor expansion versus the size parameter,
$x=2\pi a/\lambda$, is usually considered.~\cite{hulst}

At optical frequencies involved in many experiments, however, only the
time average of the electromagnetic force is observed.  In this
letter, we establish the form of the time averaged total force on a
particle without restriction on the spatial dependence of the
electromagnetic field. Further, we discuss some of its
consequences. For time harmonic electromagnetic waves,~\cite{born} we
write $\ve{E}(\ve{r},t)=\ve{E}_0 e^{-i\omega t}$,
$\ve{B}(\ve{r},t)=\ve{B}_0 e^{-i\omega t}$ and
$\ve{p}(\ve{r},t)=\ve{p}_0 e^{-i\omega t}$, $\ve{E}_0$, $\ve{B}_0$,
and $\ve{p}_0$ being complex functions of position in space, and $\Re
e$ denoting the real part. Then, the time average of the total force
is: \be
\label{averagef}<\ve{F}>  =  \frac{1}{4T}\int_{-T/2}^{T/2} \bigg[
(\ve{p}+\ve{p}^*).\ve{\nabla}(\ve{E}+\ve{E}^*)\nonumber\\
+\frac{1}{c}(\dot{\ve{p}}
+\dot{\ve{p}}^*)\times(\ve{B}+\ve{B}^*)\bigg] dt,\ee where $*$ stands
for the complex conjugate. On performing the integral and using
$\ve{E}_0$, $\ve{B}_0$, and $\ve{p}_0$, Eq.~(\ref{averagef}) yields
for each $i^{th}$ Cartesian component of the averaged total
force:\be\label{lorentzm} <F^i> = (1/2)\Re e\left[ {p_0}_j\partial^j
({E_0}^i)^*+ \frac{1}{c}\varepsilon^{ijk} \dot{p_0}_j
({B_0}_k)^*\right]\ee for $(i=1,2,3)$, where $\varepsilon_{ijk}$ is
the Levi-Civita tensor. On using the relations
$\ve{B}_0=\frac{c}{iw}\ve{\nabla}\times\ve{E}_0$,
$\ve{p}_0=\alpha\ve{E}_0$, and $\dot{\ve{p}_0}=-i\omega\ve{p}_0$ one
gets for Eq.~(\ref{lorentzm}): \be\label{lorentzm2} <F^i>=(1/2)\Re
e\bigg[\alpha \big( {E_0}_j\partial^j ({E_0}^i)^* \nonumber\\
+\varepsilon^{ijk} \varepsilon_{klm}{E_0}_j \partial^l({E_0}^m)^*
\big)\bigg].\ee On taking into account that: $\varepsilon^{ijk}
\varepsilon_{klm}=\delta^i_l\delta^j_m-\delta^i_m\delta^j_l$ one can
finally express $<F^i>$ as: \be\label{force} <F^i>=(1/2)\Re e
\left[\alpha {E_0}_j\partial^i({E_0}^j)^*\right].  \ee
Eq.~(\ref{force}) is the main result of this letter. It represents the
total averaged force exerted by an arbitrary time harmonic
electromagnetic field on a small particle.

In this connection, Ref.~[\ref{drainew}] establishes the average force
on an object represented by a set of dipoles when the electromagnetic
field is a plane wave. We notice that in this case Eq.~(\ref{force})
reduces to just Eq.~(\ref{fscatt}), in agreement with
Ref.~[\ref{drainew}]. However, as we shall illustrate next,
Eq.~(\ref{force}) permits to apply the couple dipole method (CDM) to
more complex configurations like that of a small particle in front of
a dielectric surface, under arbitrary illumination (see
Ref.~[\ref{chaumet}] for a discussion on the CDM for large
particles). Also, the absence of the magnetic field $\ve{B}_0$ in
Eq.~(\ref{force}) eases the computations.

Conversely, when Eq.~(\ref{polacm}) for the polarizability is
introduced into Eq.~(\ref{force}), one obtains for the $i^{th}$
component of the time averaged optical force: \be\label{forcefaux}
<F^i> & = & (1/2)\alpha_0\Re e
\left[{E_0}_j\partial^i(E_0^j)^*\right]\nonumber\\ & = &
(1/4)\alpha_0\Re e\left[\partial^i|\ve{E}_0|^2\right]=(1/4)
\alpha_0(\partial^i|\ve{E}_0|^2) \ee which is just the gradient
force. Notice the factor (1/4) (see e.g. Ref.~[\ref{harada}]) instead
of which the factor (1/2) for non-averaged fields often appears in the
literature (see for example
Refs.~[\ref{ashkin1},\ref{ashkin2},\ref{ashkin3}]). In agreement with
the remarks of Ref.~[\ref{draine}], now the scattering force,
Eq.~(\ref{fscatt}), vanishes and thus, $<\ve{F}>$ reduces to the
gradient force. Therefore, $\alpha_0$ must be replaced from its static
expression~(\ref{polacm}) by the addition of a damping term. This was
done by Draine,\cite{draine} who with the help of the optical theorem,
obtained: \be\label{alphacr} \alpha = \alpha_0/(1-(2/3)i k^3
\alpha_0). \ee The existence of the imaginary term for $\alpha$ in
Eq.~(\ref{alphacr}) is essential to derive the correct value for the
averaged total force due to a time varying field.

As an illustration, let the field that illuminates the particle be the
beam whose electric vector is: \be\label{champ}
E_x=\exp(-x^2/2)\exp(i(kz-\omega t)),\hspace{1mm} E_y=0,\hspace{1mm}
E_z=0 \ee On using Eqs.~(\ref{polacm}) and (\ref{champ}) in
Eq.~(\ref{force}), we find:
\begin{subeqnarray}\label{forcesplit}<F_x> &
= & -(\alpha_0/2)x \exp(-x^2)\slabel{forcesplita}\\ <F_z> & = &
0.\slabel{forcesplitb}\end{subeqnarray}

On the other hand, if the correct polarizability, Eq.~(\ref{alphacr}),
is introduced with Eq.~(\ref{champ}) into Eq.~(\ref{force}), the total
force is then expressed as:
\begin{subeqnarray}
\label{forcesplit2}<{F_x}> & = & (1/2)\Re e\left[-\alpha x\exp(-x^2)\right]
\nonumber\\ & = & \frac{-(\alpha_0/2)x\exp(-x^2)}{1+(4/9)
k^6\alpha_0^2}\slabel{forcesplit2a} \\ <{F_z}> & = &(1/2)k\exp(-x^2)
\Re e \left[-i\alpha\right]\nonumber\\ & = &
\frac{\exp(-x^2)k^4\alpha_0^2/3}{1+(4/9)
k^6\alpha_0^2}.\slabel{forcesplit2b}\end{subeqnarray} For a particle
with a radius $a\ll\lambda$, {\it e.g.} $a=10$nm, at wavelength
$\lambda=632.8$nm and $\epsilon=2.25$, the factor
$(1+(4/9)k^6\alpha_0^2)$ is very close to one (notice in passing that
the expression used in Ref.~[\ref{klumme}] for $\alpha$ makes this
factor unity). We thus see that in contrast to
Eqs.~(\ref{forcesplit}), the correct form for the polarizability,
Eq.~(\ref{alphacr}), leads to a total force given by
Eqs.~(\ref{forcesplit2a}) and (\ref{forcesplit2b}), which can be
associated to the gradient and scattering components, namely, to the
time average of Eq.~(\ref{forcem}) and Eq.~(\ref{fscatt}) with
$C_{abs}=0$, respectively.

In the case of an absorbing sphere, the dielectric constant becomes
complex and so is $\alpha_0$. Then, Eqs.~(\ref{forcesplit2}) with
$a\ll \lambda$ become:
\begin{subeqnarray} \label{fabsor}<{F_x}> & = & -(1/2)\Re 
e\left[\alpha_0\right]x\exp(-x^2) \slabel{fabsor1}\\ <{F_z}> & = &
\frac{\exp(-x^2)k^4|\alpha_0|^2}{3}\nonumber\\ & + &
\frac{k\exp(-x^2)}{2}\Im m\left[\alpha_0\right].\slabel{fabsor2}
\end{subeqnarray} The imaginary part of $\alpha_0$ does
not contribute to the component $<{F_x}>$, namely, to the gradient
force Eq.~(\ref{fabsor1}). On the other hand the absorbing and
scattering force, Eq.~(\ref{fabsor2}), exactly coincides with the
expression obtained from Eq.~(\ref{fscatt}).

\begin{figure}[H]
\begin{center}
\includegraphics*[draft=false,width=85mm]{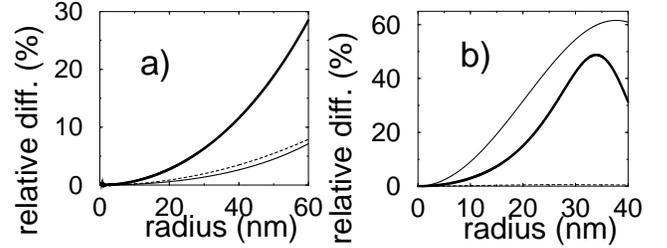}
\end{center}
\caption{ a) Relative difference between the force computed by the
exact Mie calculation and by the dipole approximation: CM-RR (full
line), LAK (thick line), DB (dashed line). The sphere is of glass
($\epsilon=2.25$), illuminated by an incident propagating plane wave
($\lambda=$600nm). b) Same as Fig.~1a for a silver sphere
($\lambda=$400nm, $\epsilon= -4+i0.7$).}
\end{figure}

\begin{figure}[H]
\begin{center}
\includegraphics*[draft=false,width=85mm]{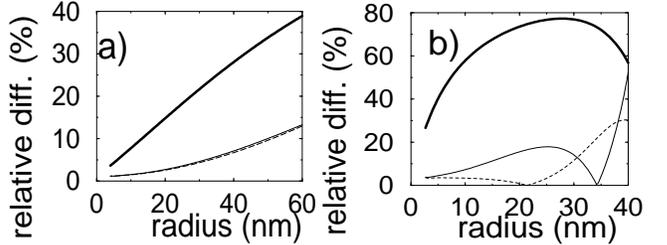}
\end{center}
\caption{ a) Relative difference between the component of the force
perpendicular to the incident wave vector obtained from CDM and by the
dipole approximation: CM-RR (full line), LAK (thick line), DB (dashed
line). The sphere is of glass ($\epsilon=2.25$) illuminated by an
incident evanescent wave ($\lambda=$600nm). b) Same as Fig.~2a for a
silver sphere ($\lambda=$400nm, $\epsilon= -4+i0.7$).}
\end{figure}

We next illustrate the above arguments with some numerical
calculations that permit us to assess the degree of accuracy of
several polarizability models previously established. We first compare
the relative difference between the force obtained from the exact Mie
calculation and the most usual polarizabilities models, namely, those
of Lakhtakia~\cite{lakhtakia} (LAK), Dungey and Bohren~\cite{dungey}
(DB), and the Clausius-Mossotti relation with the radiative reaction
term (CM-RR)~\cite{draine} previously discussed, versus the radius $a$
of a sphere illuminated by a plane propagating wave in free space
(Fig.~1a and 1b). Secondly, when this sphere is illuminated by an
evanescent wave created by total internal reflection on a dielectric
surface, the component of the force perpendicular to the incident wave
vector (Figs.~2a and 2b) is compared with the result derived from the
CDM (as discussed in Ref.~[\ref{chaumet}]). All curves are represented
up to $a=\lambda/10$. The relative difference (\%) plotted is defined
as: $100\times (F_{ref}-F_{pol.})/F_{ref}$ where $pol$ denotes the
force obtained from the corresponding method used for the
polarizability (among LAK, DB, CM-RR) and $ref$ stands for the force
derived from the Mie calculation when the incident wave is
propagating, and from the CDM when the incident wave is evanescent.

We first consider a dielectric sphere (glass, $\epsilon=2.25$)
illuminated at $\lambda=600nm$ (Figs.~1a and 2a). We observe that, for
an incident propagating wave (Fig.~1a), the result from the CM-RR
relation is better than that of DB, and this, in turn, is better than
the result from LAK. The force over a dielectric particle given by the
exact Mie calculation is:
$F=C_{scat}(1-\overline{\cos\theta})|E|^2/(8\pi)$, and that obtained
from the dipole approximation is: $F=(1/2)|E|^2\Re e[-i\alpha]$. When
the DB model is used, then $\alpha=(3/2)i a_1/k^3$ where $a_1$ is the
first Mie coefficient, hence, $4\pi \Re e[-i\alpha]$ is the scattering
cross section for an electric dipole. However, when
Eq.~(\ref{alphacr}) for the CM-RR is employed, $4\pi \Re e[-i\alpha]$
constitutes only the first term of the Taylor expansion of the
scattering cross section versus the size parameter $x$. This is why
$C_{scat}$ is underestimated when it is calculated from the CM-RR
model.Therefore, the DB model should be better. However, in both cases
the factor $\overline{\cos\theta}$ has not been taken into account in
the dipole approximation and, thus, both results overestimate the
force.  Hence, this factor $\overline{\cos\theta}$ makes a balance
making the CM-RR result closer to the Mie's solution. In the case of
an incident evanescent wave (Fig.~2a), DB and CM-RR results are very
close together, this is due to the fact that the real parts of both
polarizabilities are very close to each other. We see that LAK result,
as with a propagating wave, is far from the correct solution.

As a second example, we consider a metallic sphere (silver)
illuminated at $\lambda=400nm$ ($\epsilon=-4+i0.7$). We now observe
that for an incident propagating wave (Fig.~1b), the DB model yields
the best result. The force can exactly be written as
$F=(C_{ext}+C_{scat}\overline{\cos\theta}) |E|^2/(8\pi)$. We notice
that now $C_{scat}\overline{\cos\theta}$ is of sixth order in $x$ in
comparison with $C_{ext}$. Since $C_{ext}\propto\Re e[a_1]$ in the
electric dipole limit, the DB formulation appears as the best
here. Also, for incident evanescent waves (Fig.~2b), DB gives the most
accurate  solution. However, for a metallic sphere, the relative
permittivity much depends on the wavelength used, hence, it is now
difficult to establish a generalization of these results. We have
checked, notwithstanding, that for a gold or silver sphere in free
space in the visible, DB is often the best.

\vspace{5mm}

In summary, we have established the average total force on a little
particle in a time harmonic varying field of arbitrary form, and thus
clarify its use in the interpretation of experiments, as well as in
some previous theoretical works. For instance, we see that
Eq.~(\ref{force}) is not just the gradient force as stated in some
previous work (see e.g.  Ref.~[\ref{visscher}]). Also, this general
expression shows the importance of the radiative reaction term in the
polarizability of the sphere put forward by other authors.  Its
derivation makes no assumptions about the surrounding environment. It
is just necessary to know both the electric field and its derivative
at the position of the sphere, and thus it allows an easy handling of
illuminating evanescent fields. An immediate important consequence is
that it permits to assess the adequacy of several polarizability
models.

Work supported by the DGICYT grant PB 98-0464 and the European Union.

\end{multicols}
\end{document}